\documentclass{ws-p8-50x6-00}
\usepackage{amsmath}
\usepackage{amssymb}
\begin{document}

\title{Investigation of Bose-Einstein Correlations in 3 jet events with the
DELPHI detector}

\author{N.~van Remortel}

\address{ Universiteit 
Antwerpen (UIA), Universiteitsplein 1, B-2610 Antwerpen, Belgium\\
E-mail: nick.vanremortel@ua.ac.be}

\author{B.~Buschbeck and F.~Mandl}

\address{Institut f\"{u}r Hochenergiephysik,
  \"{O}sterr. Akad. d. Wissensch., Nikolsdorfergasse 18,  AT-1050 Vienna,
  Austria \\
E-mail: BRIGITTE@qhepu3.oeaw.ac.at, MANDL@qhepu3.oeaw.ac.at}  


\maketitle

\abstracts{
A preliminary investigation of Bose-Einstein correlations in 3 jet
events has been made by analysing the collected data at the $Z^0$ peak from
'94 and '95 and the calibration runs during the LEP2 period from '97 to
2000.
Three methods were used to extract two-particle correlation functions. No
significant difference was found between quark and gluon jets for all three
methods.}  
\section{Introduction}
\begin{figure}[t]
\begin{center}
  \begin{tabular}{c}
    \hspace{-1.cm} 
      \epsfxsize=6.0 cm 
      \epsfbox{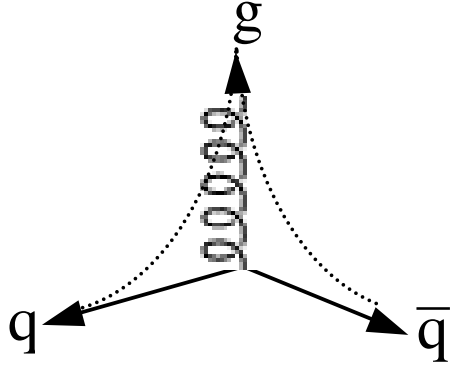} 
  \end{tabular}
\end{center}
  \caption{Lund string picture of a 3 jet $Z^0$ decay.}
  \label{scetch} 
\end{figure} 
Bose-Einstein Correlations (BEC) have been extensively investigated in $e^+e^-$
annihilations at LEP \cite{lep1adlo}. 
In the classical approach 
 \cite{hbt,gglp}
they are viewed as interference of identical bosons
 produced incoherently from their source. The source extensions can be deduced
from the momentum-difference spectra. Both ingredients: incoherence and 
symmetrization of the wave function are needed in their derivation.
In a more recent development, the Lund string picture \cite{lundbook} offers an alternative approach
for $e^+e^-$ reactions. 
It has been shown in \cite{andring} that in this coherent scenario, by making a minimum of
assumptions and without introducing extra hadronisation parameters, the  
predicted source sizes of particle production 
are typically of the order of 1 fm, in agreement with
most experimental observations. 
Both scenarios differ in predicting 
whether BEC between particles coming from different W bosons in 
$e^+e^- \rightarrow W^+W^-$ events is possible.
While in the classical approach they are unavoidably expected,
in the Lund string picture they are predicted to be absent if there
is no color (re)connection between the strings. Much effort has been
spent to test and distinguish between these two predictions
experimentally 
 \cite{lep2}.
 However, since the measurements are statistically very
limited, probably no strong statement can be made.

According to the Lund string picture there exists another reaction where 
two strings (color flux tubes) are produced together, namely in 3-jet events
 $e^+e^- \rightarrow Z^0 \rightarrow q\bar{q}g$ 
(see sketch in Fig.~\ref{scetch}). If the two strings (dotted lines in Fig.~\ref{scetch})
hadronise independently without color connection, BEC are expected to be weakened
because there will be no correlations between particles stemming from
different strings - in analogy to the situation with W-pair production \cite{eddinote}.
This is again at variance with the classical picture. In both approaches
not only the radii but also the strengths measured in gluon and quark jets could be different.

In this study we will investigate whether BEC manifests itself differently in
gluon jets and quark jets. In case of an incoherent manifestation of
inter-string BEC, an extra component with bigger radius should be observed in
gluon jets. About a possible manifestation of a coherent type of inter-string
BEC, no statements can be made.
For this study the 1994 and 1995 LEP1 data set taken by the DELPHI detector was
analysed, together with the calibration runs taken in the years 1997 till
2000, corresponding to a 3 jet event sample of respectively 236489 and 38166
selected events.
Our Monte Carlo reference samples without BEC were statistically limited and
corresponded to respectively 130244 and 69189 events for the LEP1 and LEP2
period. 
\section{Correlation Functions}
\label{sect2}
\begin{figure}[t]
\begin{minipage}{.46 \linewidth}
    \hspace{-.8cm}
    \centering\epsfig{figure=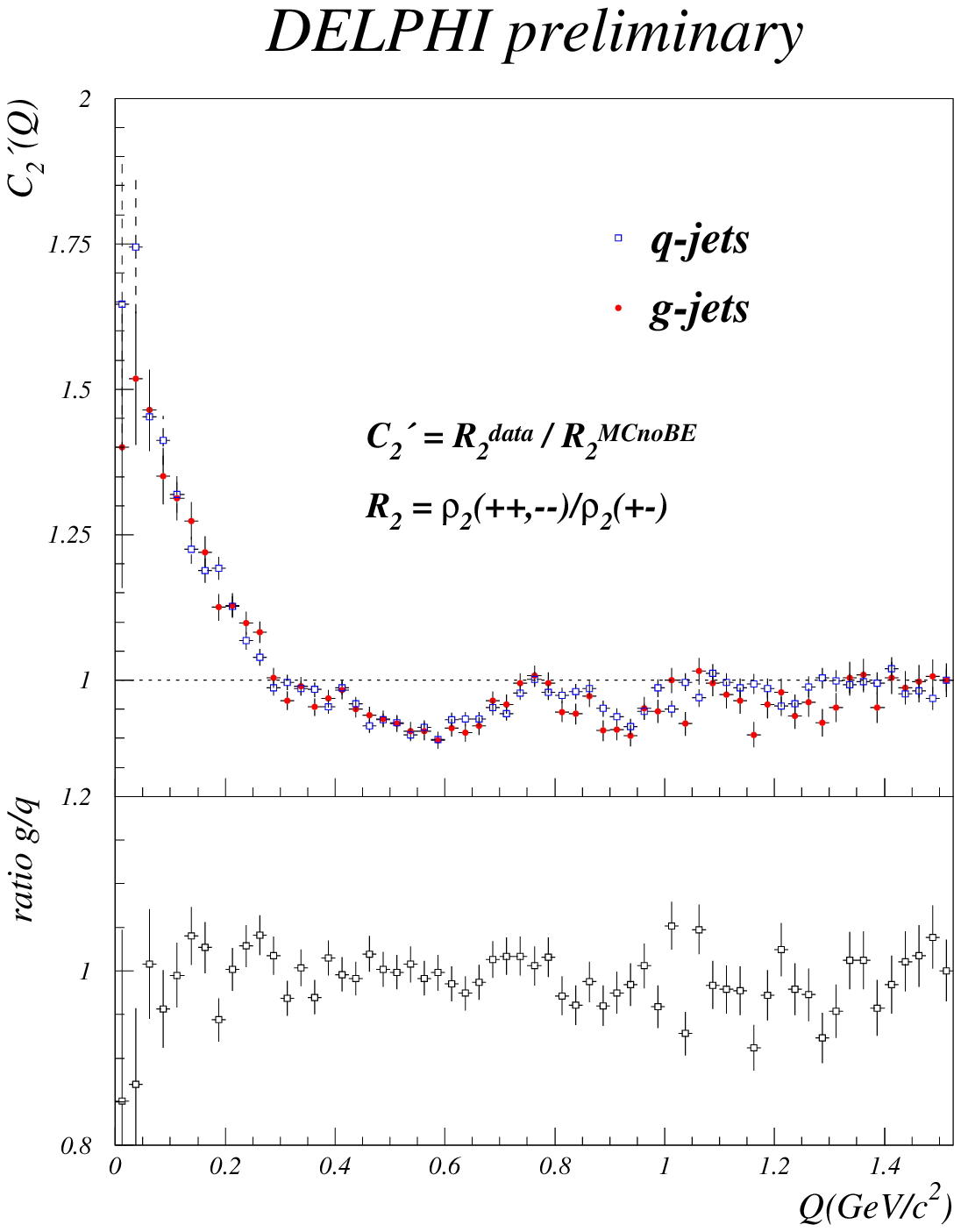,width=6.cm}
    \caption{Comparison of the $C~'_2(Q)$ for the LEP1 datasample between
    gluon jets and light quark jets. The ratio between gluon and quark jets
    is shown below. } \label{fig1}
\end{minipage}\hfill
\begin{minipage}{.46 \linewidth}
    \hspace{-1.cm}
    \centering\epsfig{figure=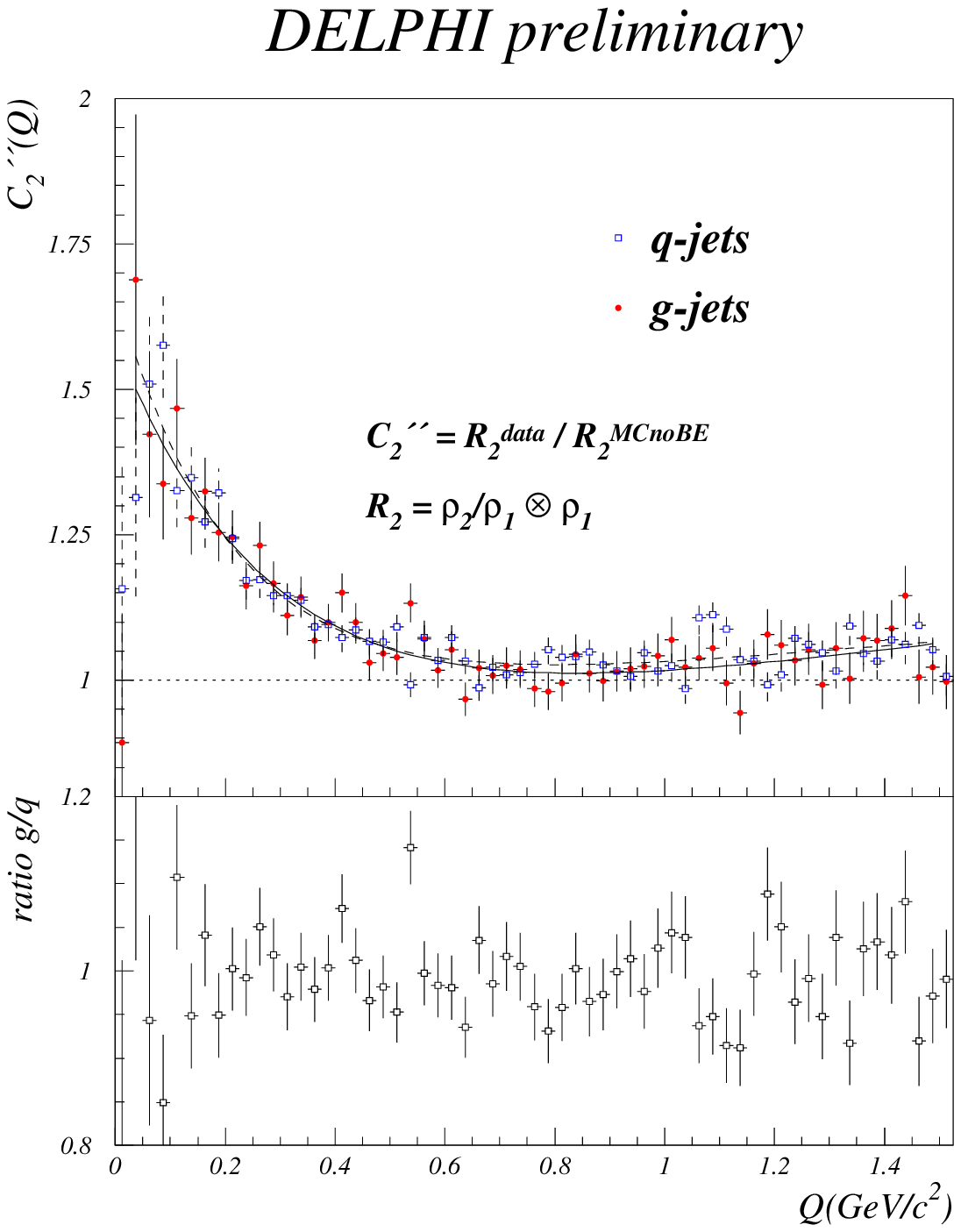,width=6.cm}
    \caption{Comparison of the $C~''_2(Q)$ for the LEP2 datasample between
    gluon jets and light quark jets. The ratio between gluon and quark jets
    is shown below.} \label{fig2}
\end{minipage}
\end{figure}
In most cases, the Bose-Einstein effect is investigated by means of
two-particle correlation functions, although there are other methods which
prove to be more accurate \cite{sarsis}.
In this note we define the two-particle correlation function as the ratio of
the two particle density of the data (or signal MC) with the two particle
density of a reference sample which does not include the Bose-Einstein effect.
The two-particle densities \cite{chekanov} are calculated as a function of the
Lorentz-invariant four momentum difference $Q=\sqrt{-(p_1-p_2)^2}$, where $p_1$
and $p_2$ are the four-momenta of the two particles:
\begin{equation}
\rho_{2}(Q)=\frac{1}{N_{ev}}\frac{dn_{\rm ~pairs}}{dQ}.
\end{equation}
Here $n_{pairs}$ stands for the number of like-sign (unlike-sign) particle
combinations. Three approaches were chosen to construct the two-particle
correlation function.
First a MC sample without any BEC was chosen to construct $C_2$:
\begin{equation}
C_2(Q)=\frac{\rho_{2}(Q)_{\rm ~signal}}{\rho_{2}(Q)_{\rm ~MC~no~BEC}}.
\end{equation}
This method is obviously the simplest, but one has to rely entirely on the
fact that the Monte Carlo reference sample has to reproduce all single
particle spectra and event shapes perfectly. 
Secondly, unlike sign particle pairs were chosen as reference, but a double
ratio with MC without BEC was chosen to correct for residual additional
correlation effects.
\begin{equation}
C~'_2(Q)=\frac{\rho_{2}(Q)_{\rm ~like-sign}/\rho_{2}(Q)_{\rm
    ~unlike-sign}}{\rho_{2}(Q)_{\rm ~MC~
no~BEC~like-sign}/\rho_{2}(Q)_{\rm~MC~no~BEC~unlike-sign}}.
\end{equation}
This approach has the advantage that data is essentially compared with data,
although reflections of resonances and detector effects, certainly for
particles close in momentum-energy space, can be different for like-sign and
unlike-sign pairs.
The last approach made use of a mixing technique, mixing particles from
several events. Again a double ratio with a MC sample without BEC was chosen
to correct for possible biases due to the mixing procedure:
\begin{equation}
C~''_2(Q)= \frac{\rho_{2}(Q)_{\rm~signal}/ \rho_{1}(Q)\otimes
  \rho_{1}(Q)}{\rho_{2}(Q)_{\rm~MC~no~BEC}/ \rho_{1}(Q)\otimes \rho_{1}(Q)_{\rm~MC~no~BEC}}.
\end{equation}
This method compares like-sign pairs from data with mixed like-sign pairs
from data, including most detector effects. One has to correct with MC however
to take into account detector resolution for close particle pairs and other
dynamical correlations.
\section{Analysis}
\begin{table}[t]
   \caption{Fit results of $C''_2(Q)$ for the LEP1 data}
  \begin{center}
\begin{tabular}{||c||c|c||}
\hline Parameter & udsc & gluon  \\
\hline  $\lambda$ & $0.685 \pm 0.046$ & $0.681 \pm 0.063$   \\
\hline  $r$ & $0.754 \pm 0.064$  & $0.697 \pm 0.067$    \\
\hline  $\delta$ & $0.126 \pm 0.027$  & $0.228 \pm 0.039$    \\
\hline  $N$ & $0.833 \pm 0.025$ & $0.752 \pm 0.030$    \\
\hline  $\chi^{2}$ & $1.5 ~(55 {\rm ~ndf})$  & $1.0 ~(55 {\rm ~ndf})$   \\
\hline \hline    
\end{tabular}
     \label{tab:01}
  \end{center}
\end{table} 
\begin{table}[t]
   \caption{Fit results of $C''_2(Q)$ for the LEP2 calibration runs}
  \begin{center}
\begin{tabular}{||c||c|c||}
\hline Parameter & udsc & gluon  \\
\hline  $\lambda$ & $0.766 \pm 0.072$ & $0.82 \pm 0.087$   \\
\hline  $r$ & $0.94 \pm 0.10$  & $0.72 \pm 0.088$    \\
\hline  $\delta$ & $0.085 \pm 0.037$  & $0.148 \pm 0.061$    \\
\hline  $N$ & $0.945 \pm 0.037$ & $0.867 \pm 0.057$    \\
\hline  $\chi^{2}$ & $1.1 ~(55 {\rm ~ndf})$  & $0.9 ~(55 {\rm ~ndf})$   \\
\hline \hline    
\end{tabular}
     \label{tab:02}
  \end{center}
\end{table} 
Since the two-particle correlation functions of
gluon jets were compared with those for jets coming from a light quark,
an anti b-tag \cite{btag} was applied to the event, reducing the b quark
contamination to 2\%. The lowest energetic jet was chosen as being the
gluon jet, while the highest energetic jet was chosen to be the quark jet
The purities of these taggings were calculated using the first order QCD
matrix element \cite{glupur}, and amounted to 78\% by requiring that the
energy of the lowest energetic jet did not exceed 15 GeV.
The comparison of the two-particle correlation function for light quark and
gluon jets was made. Two examples are shown in Figs.~\ref{fig1} and~\ref{fig2}.
Fig.~\ref{fig1} shows the comparison of the correlation functions
$C~'_2(Q)$ for the LEP1 dataset, using unlike-sign combinations as a
reference sample. Fig.~\ref{fig2} shows the comparison of the correlation
functions $C~''_2(Q)$ for the LEP2 dataset, using mixed tracks as a reference
sample. All three methods as described in section~\ref{sect2} were used for
both datasets. All comparisons between light quark and gluon jet correlation
functions showed no excess at low Q values in gluon jets wrt. quark jets,
which would indicate an extra component with bigger radius.
Finally the $C''_2(Q)$ distributions were parametrised in the region 0.025 $GeV/c^2$
$<$ Q $<$ 1.5 $GeV/c^2$, 
with an exponential function and a long range correlation term:
\begin{equation}
C~''_2(Q)=N(1+\lambda e^{-r Q})(1+\delta Q)
\end{equation}
The results of the fit are summarized in Table~\ref{tab:01} and Table~\ref{tab:02}. All errors are statistical
only and are not corrected for bin-to-bin correlations. Again, the values of
$\lambda$ and $r$ indicating the strength and the radius of the correlation
source for both datasets are compatible with each other within their errors.
\section{Conclusions}
Differences in the two-particle correlation functions were investigated in
large event samples, using the data collected in '94 and '95 and during the
calibration runs of the LEP2 period. Using 3 different methods, no extra
component was found in the two-particle correlation function for gluon jets
wrt. light quark jets. All results are preliminary.
We would like to thank E. De Wolf, G. Gustafsson, K. Hamacher and M. Siebel
for useful hints and discussions.  

\end{document}